\documentclass[conference]{IEEEtran}
\IEEEoverridecommandlockouts

\usepackage{cite}
\usepackage{amsmath,amssymb,amsfonts}
\usepackage{algorithmic}
\usepackage{graphicx}
\usepackage{textcomp}
\usepackage{xcolor}
\def\BibTeX{{\rm B\kern-.05em{\sc i\kern-.025em b}\kern-.08em
    T\kern-.1667em\lower.7ex\hbox{E}\kern-.125emX}}
    
\usepackage{booktabs}
\usepackage{acronym}
\usepackage{array}
\newcommand{\PreserveBackslash}[1]{\let\temp=\\#1\let\\=\temp}
\newcolumntype{C}[1]{>{\PreserveBackslash\centering}p{#1}}

\acrodef{BEC}[BEC]{binary erasure channel}
\acrodef{CSI}[CSI]{channel state information}
\acrodef{DMC}[DMC]{discrete memoryless channel}
\acrodef{LDPC}[LDPC]{low-density parity-check}
\acrodef{SNR}[SNR]{signal-to-noise ratio}

\begin{document}

\title{Physical-Layer Security: \\ Does it Work in a Real Environment?}

\author{\IEEEauthorblockN{Benjamin Jensen, Bradford Clark, Dakota Flanary, Kalin Norman, Michael Rice, Willie K. Harrison}
\IEEEauthorblockA{Department of Electrical and Computer Engineering,
Brigham Young University,
Provo, Utah, USA \\
\{benj4jen, brclar, dakotaf, kalinn, mdr, willie.harrison\}@byu.edu }
}

\maketitle

\begin{abstract}
This paper applies channel sounding measurements to enable physical-layer security coding. The channel measurements were acquired in an indoor environment and used to assess the secrecy capacity as a function of physical location. A variety of Reed-Muller wiretap codes were applied to the channel measurements to determine the most effective code for the environment. The results suggest that deploying physical-layer security coding is a three-point design process, where channel sounding data guides 1) the physical placement of the antennas, 2) the power settings of the transmitter, and 3) the selection of wiretap coding.  
\end{abstract}

\begin{IEEEkeywords}
physical-layer security, channel sounding, wiretap coding, Reed-Muller codes
\end{IEEEkeywords}

\section{Introduction}
As wireless transmissions grow more frequent, there is a corresponding increase in the amount of sensitive information exchanged over the air. Consequently, the demand for secure and reliable data transfer is ever expanding. Cryptography is the traditional method used to provide data confidentiality, and it relies on the assumed hardness of mathematical inverse problems such as the discrete logarithm problem~\cite{Diffie1976} or factoring large integers~\cite{Stinson_CryptoBook}. Since the number of possible keys is much smaller than the number of possible messages, a brute-force attack can defeat any modern cryptographic algorithm, even though it may take decades, or longer, to recover the message~\cite{Shannon1949}. Future advances in mathematics and computation may render a subset of these cryptographic algorithms obsolete~\cite{Shor1994}.

The field of physical-layer security~\cite{Wyner1975,BlochBook} has attracted much attention of late due to the strong information theoretic security guarantees that are possible to be made~\cite{Wyner1975, Csiszar1978, Maurer2000}. Approaches to physical-layer security are rooted in both signaling~\cite{Mukherjee2014} and coding\footnote{Codes of this type are referred to as wiretap, secrecy, or physical-layer security codes.} techniques~\cite{BlochBook, Harrison2013}, and the extra security from the physical layer can be added to security measures already in place at other layers, such as cryptography at the application layer~\cite{Harrison2009}.

Although physical-layer security has many advantages, much of the theoretical work suffers from assumptions that are regrettably impractical. This fact has slowed both the development of proof-of-concept prototypes and the acceptance of physical-layer approaches to security in practice~\cite{Johnson2018}.  Implementations to date either focus on the distribution of secret keys~\cite{PierrotTestbed2013}, or consider techniques that fall short of providing information theoretic security, albeit while making reasonable practical assumptions~\cite{MartinsTestbed2018}. One of the chief limiting assumptions made in almost all theoretical work that makes achieving information theoretic security difficult in practice is that the designer must know the \ac{CSI}, or channel parameters, for both legitimate receivers and eavesdroppers. While measuring an eavesdropper's channel parameters is impossible for some applications, there are cases where the channel can be learned through channel sounding.

In this paper, we present a methodology for deploying physical-layer security in a real-world wireless communication system. The method requires thorough channel sounding within a controlled environment so as to characterize the capacity for physical-layer security as a function of the placement of both legitimate antennas and eavesdroppers. Using the channel sounding measurements, we then show how to adjust both the power at the transmitter along with the choice of coding so as to maximize the reliable throughput to a legitimate receiver without sacrificing security.

The remainder of the paper is organized as follows. Section~\ref{sec:setup} gives the system setup, including the wiretap channel model and the basic coding approach to physical-layer security. Channel sounding procedures are explained in Section~\ref{sec:channel_sounding} and basic channel measurements are shown for an indoor test environment. Section~\ref{sec:coding} presents the secrecy capacity analysis as a function of physical location of the eavesdropper, and shows how specific wiretap codes can be deployed to achieve reasonable rates while providing tandem reliability and security in the test network. Finally, conclusions and a short outline of future work are given in Section~\ref{sec:conclusion}.

\section{System Setup} 
\label{sec:setup}

We first provide a quick guide to the notation of the paper. Capital letters represent either random variables or matrices, which is made clear by context. The lengths of random vectors are defined by superscripts on the random variables, except where the notation is cumbersome. Then the size is given explicitly in the text. The $i$th random variable in a random vector is labeled with subscript $i$.  Realizations of random variables are given by their lowercase equivalents, often without including superscripts for vectors, and constants are also written as lowercase variables. We also let $\mathbb{F}_2$ indicate the binary field of numbers, and indicate the size of binary vector spaces over $\mathbb{F}_2$ (i.e., vectors or matrices) with superscripts.

\subsection{Wiretap Channel Model}
\label{subsec:WTC}

\begin{figure}
  \includegraphics[width=\linewidth]{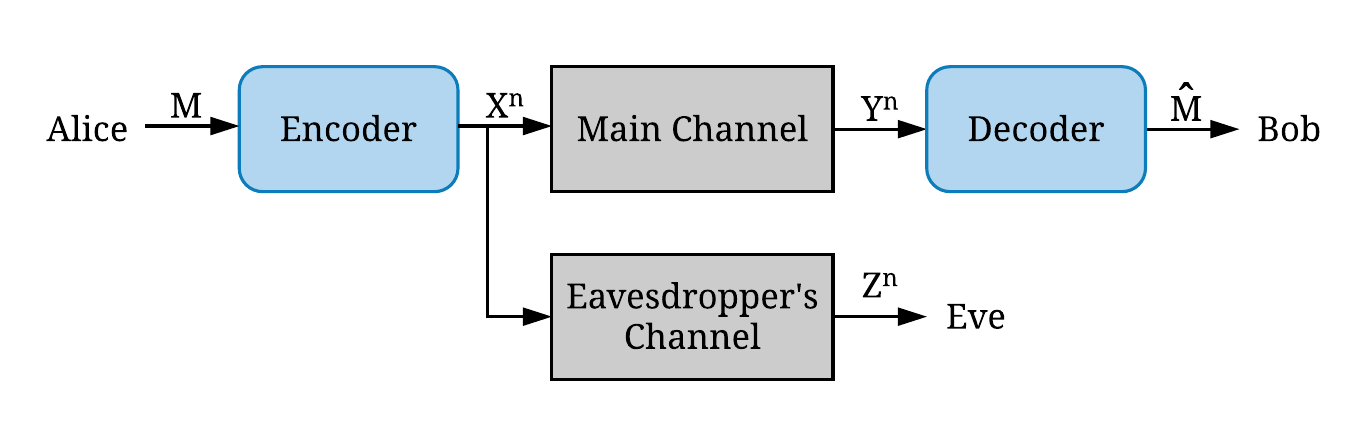}
  \caption{The wiretap channel model.}
  \label{fig:WiretapChannel}
\end{figure}

The wiretap channel, developed by Wyner in 1975~\cite{Wyner1975}, provides a simple model for wireless communication and assumes three users: Alice, the transmitter; Bob, the intended recipient; and Eve, a passive eavesdropper. A modern version of this model is given in Fig.~\ref{fig:WiretapChannel}. The figure illustrates that Alice encodes a message $M$ into a codeword $X^n$, which is then transmitted to Bob over the \emph{main channel} of communications. Bob receives $Y^n$, which may be different from  $X^n$ because of the noise in the channel. He then decodes $Y^n$ and gets $\hat{M}$, an estimate of $M$. Eve also receives a noisy version of $X^n$, albeit through a separate channel called the \emph{eavesdropper's channel}. Because Eve's physical position is different from Bob's, her observation $Z^n$ is assumed to be different from Bob's observation $Y^n$. Thus, Eve faces a different challenge in retrieving the message $M$, and Alice and Bob design their encoder to exploit this difference for security.

Since this paper focuses on real-world channel measurements, it may be reasonable to assume that we would consider the case where both the main and wiretap channels are Gaussian~\cite{Leung-Yan-Cheong1978}, and we will certainly do this with reference to the secrecy capacity in Section~\ref{sec:coding}. However, coding for information theoretic security over the Gaussian wiretap channel is still an open problem.  Although it has been shown to be possible to code over the Gaussian wiretap channel model and achieve information theoretic security and reliability in tandem~\cite{Ling2014}, the only explicit coding techniques designed for this channel either address security in terms of error rate rather than information theory~\cite{Klinc2011, Baldi2012}, or attempt to add additional layers of coding so that the Gaussian wiretap channel mimics a \ac{DMC} model instead~\cite{Sarmento2015, Harrison2018}. 

For this initial study, we design codes for a discrete memoryless variant of the wiretap channel model, where the main channel is assumed to be error-free, and the eavesdropper's channel is a \ac{BEC}. We justify this model by noting that several communication systems tend to give extremely reliable estimates of transmitted data or seemingly no information whatsoever. Some examples are packet-based systems with inter-packet interleaving~\cite{Harrison2011}, and systems that deploy modern error-control coding such as \ac{LDPC} codes or turbo codes~\cite{MoonCoding}. To be more specific, we assume a threshold $\tau$ in \ac{SNR} over a Gaussian link, such that
\begin{equation}
\label{eq:modelBob}
    y = \begin{cases} x & \mbox{if $\text{\ac{SNR}}_m \geq \tau$} \\ ? & \mbox{if $\text{\ac{SNR}}_m < \tau$}, \end{cases}
\end{equation}
and 
\begin{equation}
\label{eq:modelEve}
    z = \begin{cases} x & \mbox{if $\text{\ac{SNR}}_e \geq \tau$} \\ ? & \mbox{if $\text{\ac{SNR}}_e < \tau$}, \end{cases}
\end{equation}
where `$?$' indicates an erasure, $\text{\ac{SNR}}_m$ is the \ac{SNR} over the main channel, and $\text{\ac{SNR}}_e$ is the \ac{SNR} over the eavesdropper's channel. The variables $x$, $y$, and $z$ are signals that depict a single use of the wiretap channel in Fig.~\ref{fig:WiretapChannel}. Since some symbols may be erased over the main channel, but the codes assume an error-free link, we will use channel sounding measurements on parallel channels to choose only highly reliable carriers for transmission to Bob. More details on this are given in Sections~\ref{sec:channel_sounding} and~\ref{sec:coding}.

The encoder is designed with two constraints in mind, namely:
\begin{enumerate}
    \item $\Pr(M\neq \hat M) < \delta_r$ (reliability constraint), and
    \item $\mathbb{I}(M;Z^n) < \delta_s$ (security constraint).
\end{enumerate}
Here 
\begin{equation}
    \mathbb{I}(M; Z^n) = \mathbb{H}(M) - \mathbb{H}(M|Z^n),
\end{equation} 
which is the usual mutual information~\cite{cover_thomas_2006}, and  $\mathbb{H}(M|Z^n)$ is called the \emph{equivocation}~\cite{Wyner1975}. It is assumed that the statistics of the message are time invariant, and thus, the entropy of the message $\mathbb{H}(M)$ is a constant. Clearly then, achieving the security constraint amounts to making $\mathbb{I}(M;Z^n)$ small, or equivalently, making the equivocation large. The constants $\delta_r$ and $\delta_s$ are chosen by the designer, and are assumed to be small. Notice that we do not take the traditional asymptotic approach~\cite{Wyner1975,Maurer2000} in the security constraint, but rather consider the exact amount of mutual information (or equivalently, equivocation) at the eavesdropper. This approach allows us to work in the finite blocklength regime, making the results more appropriate for deployment of physical-layer security systems. 

\subsection{Coding for Secrecy}

For this paper, messages are assumed to be chosen uniformly at random from an alphabet $\mathcal{M} = \{1,2,\ldots,2^k\}$, and converted to binary messages of $k$ bits each. All codes considered are binary with blocklength denoted by $n$. Thus, the rate of a wiretap code is $R = k/n$. We consider only some of the simpler wiretap code structures in this initial work, i.e., coset coding at finite blocklength~\cite{Wyner1975, Harrison2018a}. Since the main channel is assumed to be error-free, we devote the entire overhead of the code to achieving the security constraint.

Consider an $(n, n-k)$ linear block code $\mathcal{C}$ with $(n-k)\times n$ generator matrix $G$ and $k\times n$ parity-check matrix $H$. Let $M'$ be an $(n-k)$-bit auxiliary message that is uniformly distributed over $\mathbb{F}_2^{n-k}$. Then the encoding function is given by
\begin{equation}
    x^n = \left[\begin{matrix} m & m' \end{matrix}\right]\left[\begin{matrix} G' \\ G \end{matrix}\right],
\end{equation}
where $G'$ is chosen from $\mathbb{F}_2^{k\times n}$ so that its rows form a full rank matrix when stacked with $G$. Oftentimes the parity check matrix $H$ can be used as $G'$. Since $m'$ is chosen uniformly at random, the mapping from $m$ to $x^n$ is a one-to-many mapping, where $x^n$ can take on any of $2^{n-k}$ codewords. The structure of the encoder, however, ensures that every possible $x^n$ for a given $m$ is in the same coset~\cite{MoonCoding} of $\mathcal{C}$, and that the cosets expand to fill the vector space $\mathbb{F}_2^n$~\cite{Wyner1975,Harrison2013,Harrison2018a}.

A simple example will suffice to illustrate this process. Let $n=4$, $k=2$,
\begin{equation}
    G = \left[ \begin{matrix} 0 & 1 & 1 & 1 \\ 1 & 1 & 1 & 0 \end{matrix} \right],
\end{equation}
and 
\begin{equation}
    G' = H = \left[ \begin{matrix} 1 & 1 & 0 & 1 \\ 1 & 0 & 1 & 1 \end{matrix} \right].    
\end{equation}
The mapping of messages to codewords is then given in Table~\ref{tab:codeExample}. Note that $\mathcal{C}$ is given in the first row of the table, and each row of the table forms a coset of $\mathcal{C}$.

\begin{table}
    \caption{An example wiretap code with block length $n = 4$ and rate $R = 1/2$.}
    \label{tab:codeExample}
    \begin{center}
    \begin{tabular}{ccccc}
    \toprule
        $m$ \textbackslash $m'$ & \textbf{00} & \textbf{01} & \textbf{10} & \textbf{11}     \\ \midrule
        \textbf{00} & 0000 & 1110 & 0111 & 1001 \\ [5pt]
        \textbf{01} & 1011 & 0101 & 1100 & 0010 \\ [5pt]
        \textbf{10} & 1101 & 0011 & 1010 & 0100 \\ [5pt]
        \textbf{11} & 0110 & 1000 & 0001 & 1111 \\ \bottomrule
    \end{tabular}
    \end{center}
\end{table}

The decoder amounts to a calculation of the syndrome $s^k$ of the transmitted codeword, i.e.,
\begin{equation}
    s^k = y^n H^T.
\end{equation}
Since the main channel is noise-free, $y^n=x^n$. There then exists a mapping from the syndrome to the message, which is a function of the selection of the matrix $G'$. In~\cite{Pfister2017}, it was shown that $G'$ could be chosen to ensure 
\begin{equation}
\label{eq:syndrome}
    m^k = s^k,
\end{equation}
although a simple look-up table could provide the mapping from $s^k$ to $m^k$ were this not to be the case. Note that for the example in Table~\ref{tab:codeExample}, the expression in (\ref{eq:syndrome}) holds.

In designing codes to satisfy the security constraint given in Section~\ref{subsec:WTC}, we will find it convenient to analyze a code in terms of its worst-case leakage given $\mu$ observed coded bits at the eavesdropper. Note that this is the coding approach given in the wiretap-II case~\cite{Wyner1984}. To aid in this endeavor, consider the equivocation matrix~\cite{Harrison2018a}. This matrix gives us a straightforward mechanism for cataloguing the number of erasure patterns with $\mu$ revealed coded bits that maintain a certain number of bits of equivocation. It was shown in~\cite{Harrison2018a} that wiretap codes based on cosets of a linear block code can only leak full bits of information, and thus the matrix form of the equivocation matrix is justified. 

In Table~\ref{tab:codeHierarchy} we show the tabular form of the equivocation matrix for the example code in Table~\ref{tab:codeExample}. Note that the worst-case equivocation when $\mu=2$ is one bit, although five out of six erasure patterns where $\mu=2$ maintain full equivocation (i.e., do not leak information). The worst case for $\mu=2$ is when the 2nd and 3rd bits of the transmitted codeword are revealed to an eavesdropper. Suppose, e.g., that $z = (?00?)$. Note in Table~\ref{tab:codeExample} that half of the cosets can be ruled out since the four consistent codewords to this observation indicate that $m$ is either $(00)$ or $(11)$. Thus, the equivocation is only a single bit of information, assuming the two remaining messages to be equally likely. Further note that any observation $z$ where the middle two bits are revealed will leak one bit of information to Eve~\cite{Harrison2018a}. Thus, we need only count the unique erasure patterns in the equivocation matrix, rather than all possible observations, and the worst-case equivocation given $\mu$ observed bits is easily identified when the equivocation matrix can be formed.

\begin{table}
    \caption{The equivocation matrix for the wiretap code shown in Table \ref{tab:codeExample}.}
    \label{tab:codeHierarchy}
    \begin{center}
    \begin{tabular}{cccccc}
    \toprule
        Equivocation &  \multicolumn{5}{c}{Number of} \\
        (bits) & \multicolumn{5}{c}{Revealed Bits ($\mu$)} \\ \midrule
          & 0 & 1 & 2 & 3 & 4 \\ \midrule
        2 & 1 & 4 & 5 & 0 & 0 \\ [5pt]
        1 & 0 & 0 & 1 & 4 & 0 \\ [5pt]
        0 & 0 & 0 & 0 & 0 & 1 \\ \bottomrule
    \end{tabular}
    \end{center}
\end{table}

Unfortunately when codes get larger, finding the exact equivocation matrix can get complicated. In essence, the number of patterns to check becomes too large to efficiently produce the entire matrix~\cite{Harrison2018a}. Luckily, for several families of algebraic codes, there exists a shortcut for finding the worst-case equivocation pattern for all possible $\mu$. In~\cite{Wei1991}, the generalized Hamming weights of the dual code $\mathcal{C}^\perp$ were shown to indicate the exact $\mu$ for which the worst-case pattern leaks an additional bit of information when compared to patterns where $\mu-1$ coded bits are observed. In other words, the worst-case leakage when $\mu$ coded bits are observed is equal to the number of generalized Hamming weights of $\mathcal{C}^\perp$ that are equal to or less than $\mu$~\cite{Harrison2018a,Wei1991}.

\section{Channel Sounding}
\label{sec:channel_sounding}
In this section, we demonstrate how it is possible to learn the eavesdropper's channel so as to allow for wiretap code design. We assume that legitimate and eavesdropper receivers both use the same receiver hardware. In practice, this indicates that one should perform the channel sounding experiments with the highest quality receiver hardware that an eavesdropper may be expected to use.

\subsection{Environment}

\begin{figure}
  \includegraphics[width=\linewidth]{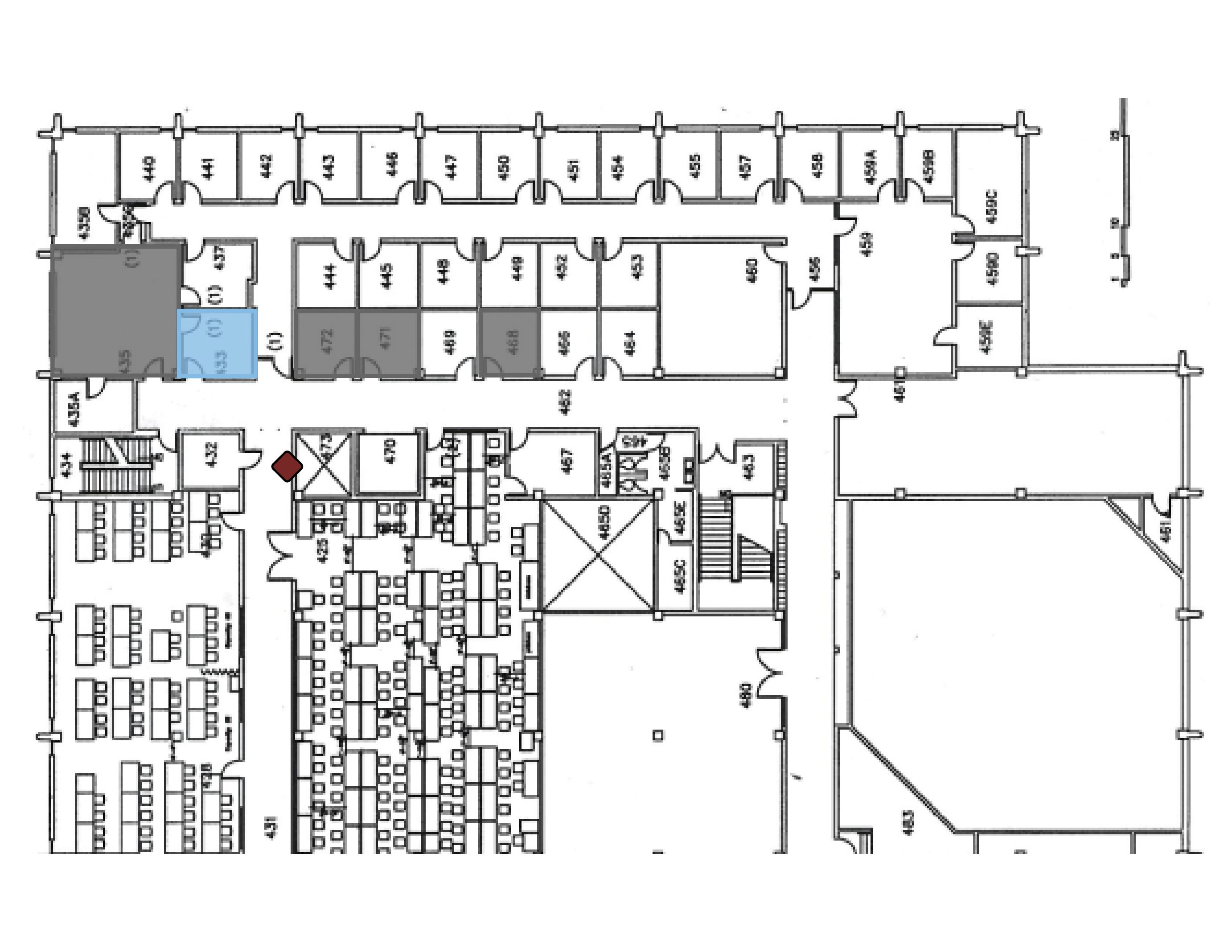}
  \caption{Indoor environment where channel sounding measurements were recorded. Grey rooms represent the potential locations of Eve while the blue room represents the location of Bob. The transmitter location is indicated by a red diamond.}
  \label{fig:preheat}
\end{figure}

The procedure was implemented in the south end of the fourth floor of the Clyde Engineering Building at Brigham Young University (see Fig. \ref{fig:preheat} for the floor map). The area in which data were gathered is comprised mostly of cinder block walls, encased in bricks, without any windows. All the data were gathered indoors. 
One office was selected as Bob's location, which is highlighted in blue in Fig.~\ref{fig:preheat}. Adjacent offices were used as potential locations for Eve, and these are highlighted in grey. The transmitter is also indicated by a red diamond.

\subsection{Procedure} 

The channel sounding signal used was motivated by 802.11g: OFDM comprising 64 sub-carriers with a sub-carrier separation of 0.3125 MHz. 
To avoid interference with 802.11g in the 2400 MHz band, the channel sounding was performed one octave below, at 1250 MHz.
The transmitter was a software-defined radio (SDR) configured to transmit 64 unmodulated sub-carriers.
A modest power amplifier was attached to the transmitter to increase the signal-to-noise ratio at the receiver.
The receiver was a second SDR that converted the received RF signal to I/Q baseband and sampled it at 20 Msamples/s.
At each location, 32 periods of the sounding signal were recorded.
The power spectral density of the channel was estimated using Welch's method of averaged periodograms \cite{hayes1996} based on length-128 FFTs.
The length of the FFT was twice the number of sub-carriers.
The even-indexed FFT bins corresponded to the sub-carriers and were used to estimate the channel gains at each sub-carrier.
The odd-indexed FFT bins were used to estimate the noise variance. 

\begin{figure}
    \centering
    \includegraphics[width=\linewidth]{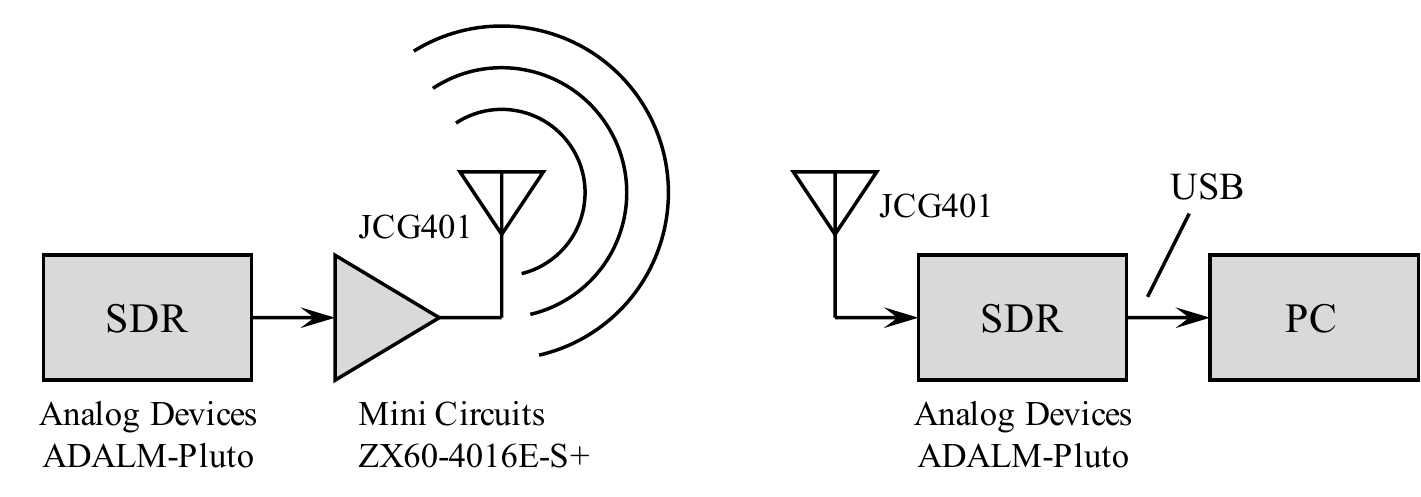}
    \caption{The system used for transmitting and receiving signal data.}
    \label{fig:channelSounding}
\end{figure}

The system for the channel sounding experiments is modeled in Fig. \ref{fig:channelSounding}, as well as the equipment used.
The transmitter was stationary at a height of 200 cm above the floor to approximate the height of a WiFi access point.
The receiver captured a ``snapshot'' of the channel on a grid with 10.2 cm spacing. 
The spacing ensured a less than half wavelength spatial sampling rate. 
The receiver height was fixed at 143 cm above the floor. 

\subsection{Basic Channel Sounding Results}

Since we performed the channel sounding with the idea of applying erasure wiretap coding to the data, we present the measurements in terms of the number of the 64 sub-carriers of the OFDM-like waveform, where a reliable signal is indicated by the model in (\ref{eq:modelBob}) and (\ref{eq:modelEve}) for each sub-carrier. Increasing the value of the threshold $\tau$ in (\ref{eq:modelBob}) and (\ref{eq:modelEve}) is equivalent to reducing the transmit power within the real environment. Fine-tuning of the $\tau$ value we apply to the measurements indicates careful selection of the transmit power so as to deliver a signal advantage to Bob over Eve. We chose $\tau$ to heuristically minimize the signal strength in each of Eve's potential locations, while maintaining a reasonable amount of signal strength in Bob's office, which resulted in  $\tau = 25$ dB for our data set. We optimize this choice of $\tau$ later. 

\begin{figure}
    \centering
    \includegraphics[width=\linewidth]{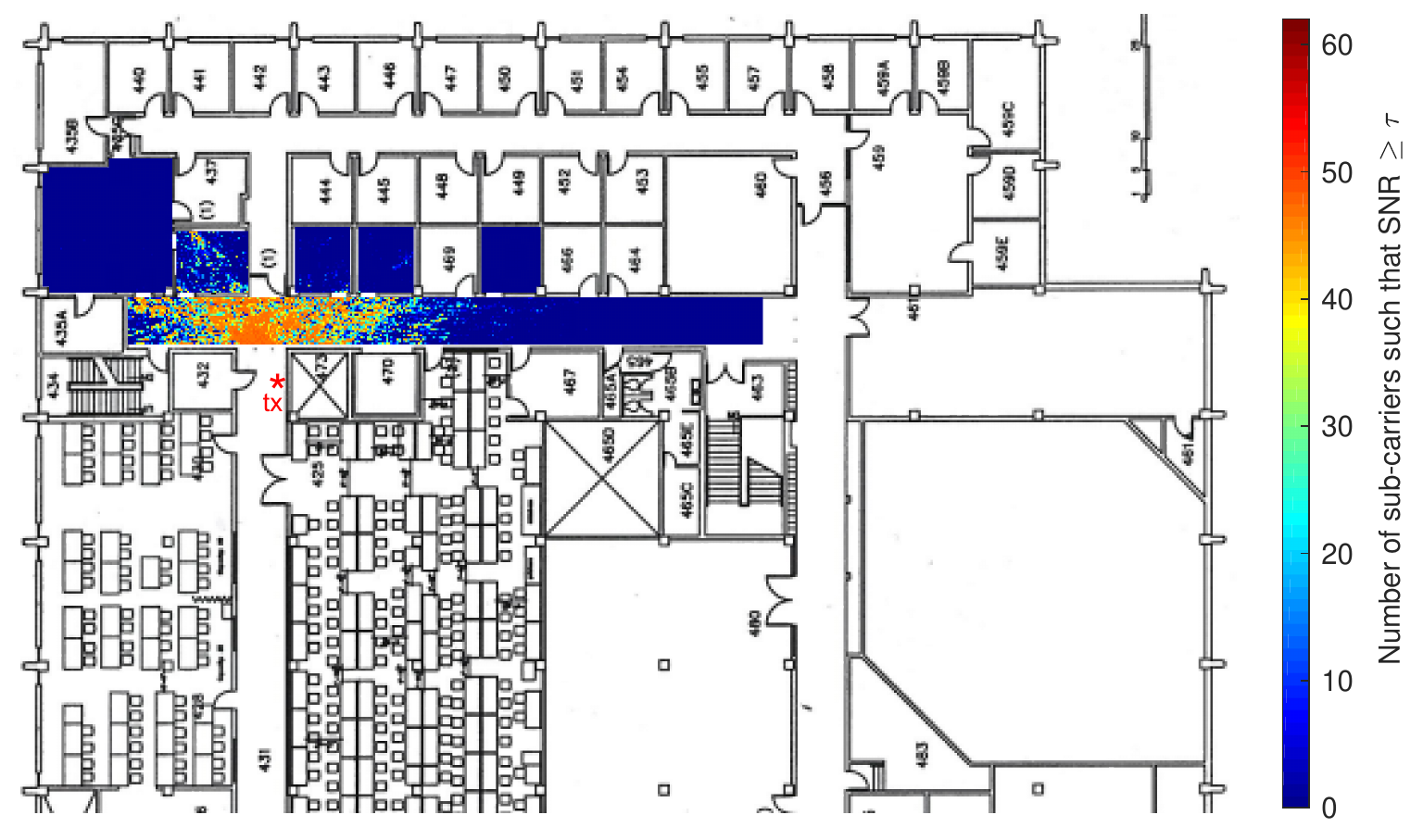}
    \caption{Heat map of the engineering building showing the number of sub-carriers such that the signal-to-noise ratio is at least 25 dB.}
    \label{fig:HeatMap}
\end{figure}

Multipath in the indoor environment proved to play a significant role in determining the utility of specific sub-carriers. The channel sounding results are illustrated in Fig.~\ref{fig:HeatMap}, where we see that direct line-of-sight also plays a significant role in the number of reliable sub-carriers, as we expect. We see that many of the hallway locations will be necessarily vulnerable to eavesdropping. In this application, however, we assume eavesdroppers will not set up their receiving antennas in the hallway, as this location could be secured through physical monitoring.

\section{Application of Coding to Match the Channel}
\label{sec:coding}

Motivated by the results in Fig.~\ref{fig:HeatMap}, we see that Bob's receiving location does maintain an advantage in signal quality over all potential eavesdropping locations. The placement of the transmit antenna was selected to deliver this advantage. In this section, we analyze the secrecy capacity of this setup, and detail the choice of wiretap codes.

\subsection{Channel Capacity}

\begin{figure}
    \centering
    \includegraphics[width=\linewidth]{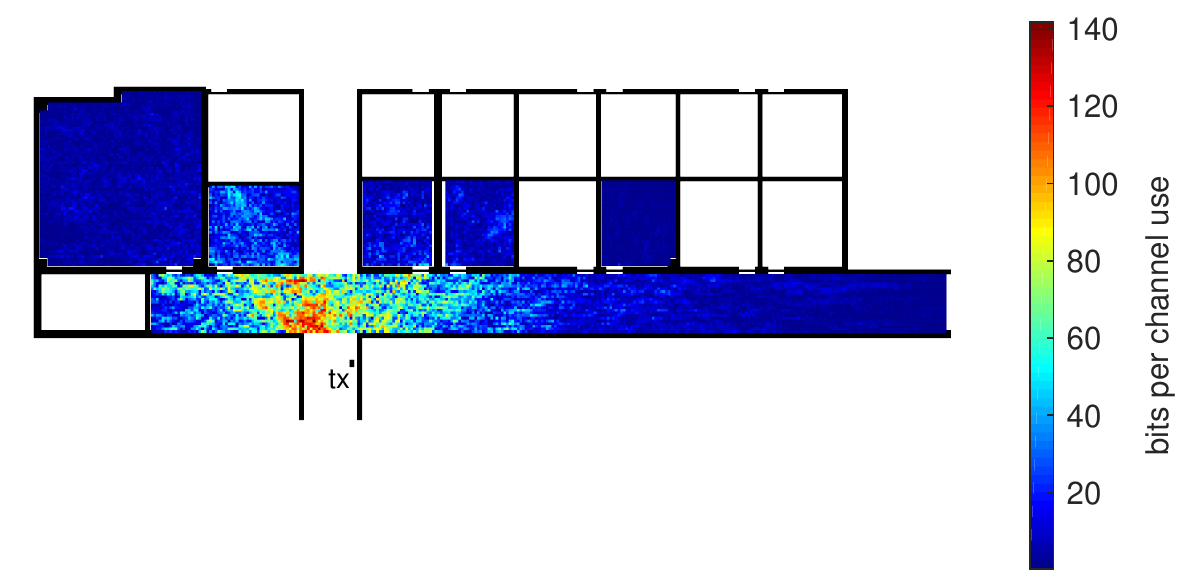}
    \caption{Channel capacity as a function of position, where `tx' indicates the location of the transmitter.}
    \label{fig:Capacity}
\end{figure}

Recall the famous Shannon result for channel capacity over a Gaussian channel~\cite{Shannon1948,cover_thomas_2006} as 
\begin{equation}
    C = \frac{1}{2} \log_2\left(1+\frac{P}{N}\right),
\end{equation}
where $\frac{P}{N}$ is the signal-to-noise ratio and $C$ is measured in bits per channel use. Assume the receiving power is given as $P_i$ for sub-channel $i \in \{1,2,\ldots,64\}$, and the noise power for all sub-channels is $N$. If the sub-channels are chosen so as not to interfere with one another, then the total capacity of the channel is~\cite{cover_thomas_2006}
\begin{equation}
\label{eq:cap}
    C = \frac{1}{2} \sum\limits_{i=1}^{64} \log_2\left(1 + \frac{P_i}{N}\right).
\end{equation}
Note that we do not discuss a power constraint here, but in this work we carefully select the transmit power to maximize secure throughput. Applying (\ref{eq:cap}) to the channel sounding measurements outlined in Section~\ref{sec:channel_sounding}, we present the capacity as a function of physical location for all testing sites in Fig.~\ref{fig:Capacity}, given our OFDM-like waveform with 64 sub-carriers. The threshold $\tau = 27$ dB was chosen to maximize the secure throughput. This figure is zoomed into the tested areas so as to better view the results, where we see that Bob's office has a greater capacity for communication than do all of the other offices. This implies that one should be able to communicate both reliably and securely in this environment.

\begin{figure}
    \centering
    \includegraphics[width=\linewidth]{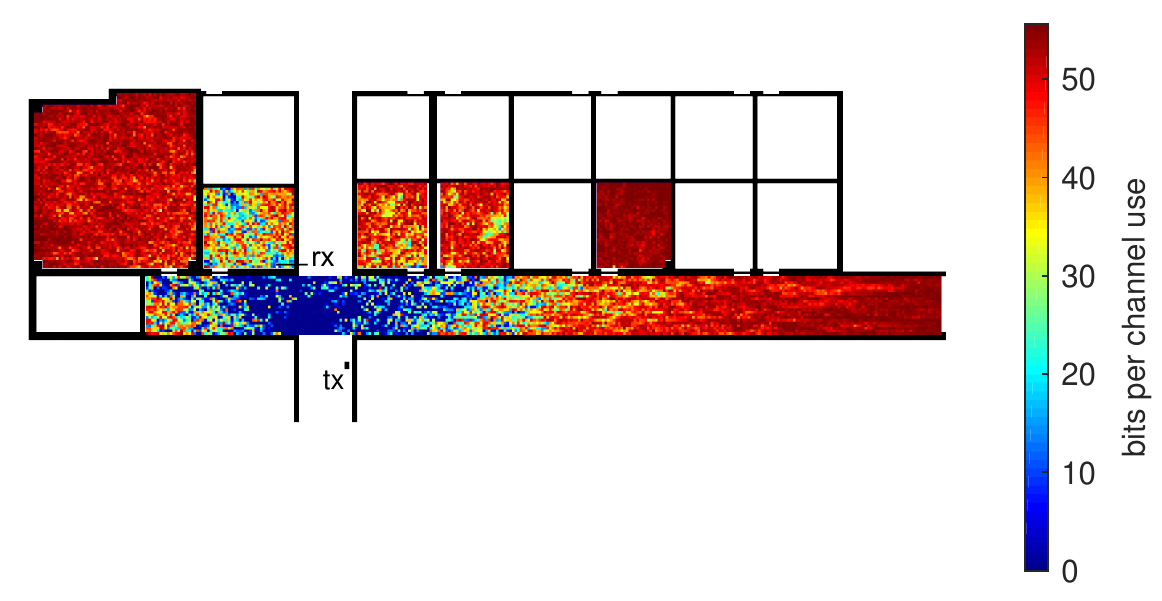}
    \caption{Secrecy capacity as a function of position, where `tx' indicates the location of the transmitter and `rx' indicates the reference location, corresponding to the best location in Bob's office.}
    \label{fig:SecrecyCapacity}
\end{figure}

To get a better feel for the advantage in capacity for Bob over Eve, we consider the secrecy capacity~\cite{Wyner1975,BlochBook}. While the capacity is the supremum of all rates at which one can communicate reliably with arbitrarily low bit error rate at the decoder~\cite{Shannon1948}, the secrecy capacity is the supremum of rates at which one can communicate while satisfying both the reliability and security constraint to arbitrarily small constants $\delta_r$ and $\delta_s$. In the analysis, $n$ is allowed to go to infinity, but we will later compare these secrecy capacity results with actual rates of codes achieved for our scenario. It is known that the secrecy capacity over the Gaussian wiretap channel~\cite{Leung-Yan-Cheong1978,BlochBook} is simply the difference of the capacity of the main channel and the capacity of the wiretap channel when the difference is positive, and zero otherwise. Thus, the overall secrecy capacity when 64 parallel sub-channels are employed is
\begin{equation}
C_s = \sum^{64}_{i=1} \max[C_{m,i} - C_{e,i}, 0],
\end{equation}
where $C_{m,i}$ is the capacity for the $i$th sub-channel as measured at the receiver in Bob's office, and $C_{e,i}$ is the capacity of the $i$th sub-channel at the eavesdropper's receiver. Fig.~\ref{fig:SecrecyCapacity} shows the secrecy capacity assuming 64 sub-channels and the channel sounding measurements with $\tau = 27$ dB. Bob's receiver is set to the location in his office that maximizes his capacity, and colors at other locations indicate $C_s$ if the eavesdropper's antenna were to be placed at that location. Notice that in the furthest offices from Bob's, the secrecy capacity is the highest, while closer offices require lower rates to achieve both reliability and security. Locations in direct line-of-sight to the transmitter have $C_s \approx 0$, as expected.

\subsection{Reed-Muller Secrecy Codes}

For the tests in this paper, we are restricted to code families for which we know the generalized Hamming weights of the dual codes, so as to allow us to identify the worst-case equivocation when the codes are used over our real-world scenario. Code families with known generalized Hamming weights given in~\cite{Wei1991} are essentially restricted to algebraic codes. The authors of~\cite{Harrison2018a} have recently shown strong evidence that both Hamming codes and their duals (simplex codes) are best for their sizes in terms of maximizing the equivocation, but the rates of these codes are quite limited. Reed-Muller (RM) codes were finally chosen for the tests presented in this paper due to the fact that the rates of length-$n$ RM codes are essentially evenly spaced between zero and one. Also, as $n$ gets larger, more RM codes exist; thus giving a wider range of rate options.

Since the duals of RM codes are RM codes~\cite{MoonCoding}, then knowing the generalized Hamming weights of RM codes is sufficient to characterize the worst-case equivocation when these codes are used as the linear block code for the coset-based wiretap code construction. (See~\cite{Wei1991} for more on the nature of the generalized Hamming weights of RM and other algebraic codes.) 

\subsection{Results}

\begin{figure}
  \centering
  \includegraphics[width=\linewidth]{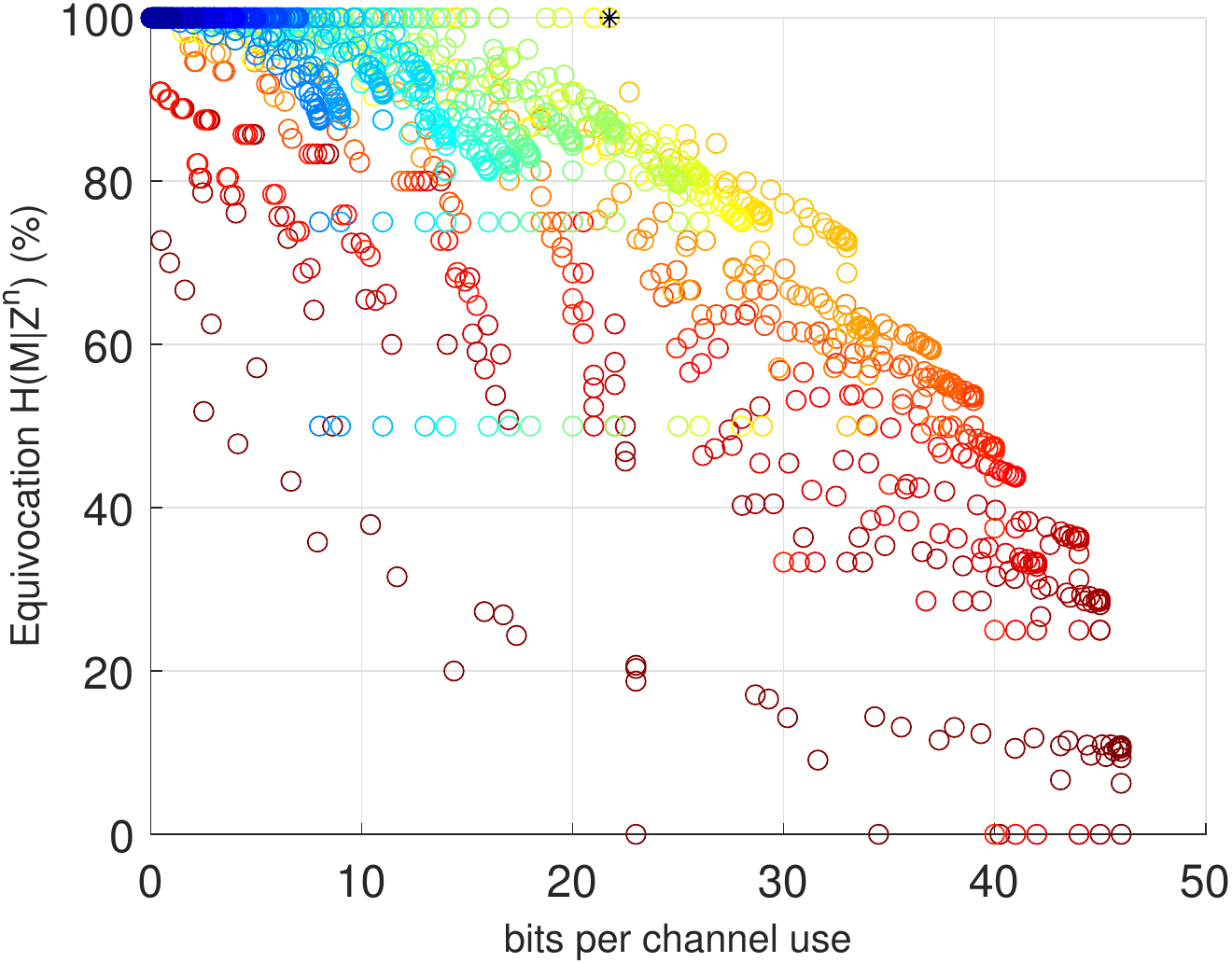}
  \caption{Graph of throughput versus percent equivocation for different RM codes and $\tau$ thresholds. Different colors signify $\tau$ values, with red indicating $\tau = 25$ dB, and blue indicating $\tau = 31$ dB. Throughput rate is calculated as the rate of the code under test multiplied by the number of sub-carriers with $SNR_m > \tau$ at Bob's receiver.} 
  \label{fig:SecurityCode}
\end{figure}

We wish to find the RM code that, when used as a wiretap code, maximizes the throughput of information subject to the constraint of full equivocation at all possible eavesdropper locations. Our results are optimistic in terms of throughput since the analysis assumes the channel sounding measurements in Fig.~\ref{fig:HeatMap} hold for all time, which is surely not true in practice. Furthermore, we assume the thresholding simplification of the Gaussian channels in (\ref{eq:modelBob}) and (\ref{eq:modelEve}), which allow us to achieve higher rates in the model than $C_s$ over the real channel. This fact implies that real-world scenarios that can be modeled like (\ref{eq:modelBob}) and (\ref{eq:modelEve}) are more capable of secrecy than the basic theory may imply. Results for a variety of codes and $\tau$ values are given in Fig.~\ref{fig:SecurityCode}. For each value of $\tau$, only the sub-carriers for which Bob has a reliable link are used. The equivocation is given in percentage so that all codes can be compared in the same figure. The rating of 100\% equivocation indicates that $\mathbb{I}(M;Z^n) = 0$. Throughput is given as the rate of the code multiplied by the number of active carriers, and is thus measured in bits per channel use. One channel use allows for all sub-carriers to be used one time. We assume BPSK modulation on each sub-carrier. 

Colors indicate choice of $\tau$, and we quickly see in the figure that fine-tuning the power at the transmitter is essential to achieving the highest possible throughput while maintaining secure communications. The range of $\tau$ values presented in Fig.~\ref{fig:SecurityCode} is 25 dB to 31 dB. Values of $\tau$ that are, respectively, higher or lower than this range indicate cases where reliable communication to Bob is not possible, or we are transmitting at such a high power level that no security can be possible using the physical layer. 

Notice that under the assumptions made in the analysis of this paper, we have several codes that are capable of maintaining secure data transfer at the 100\% level. Assuming a need for complete secrecy, the best code for our scenario on the 4th floor of the Clyde Engineering Building is the RM($u=1, m=2$) code, where $u$ indicates the order and $m$ indicates the degree of the RM code~\cite{MoonCoding}. The code is marked by a black star in Fig.~\ref{fig:SecurityCode}, and it achieves 21.75 bits per channel use of secure throughput when $\tau = 27$ dB. The minimum secrecy capacity over all possible eavesdropper locations is only 8.071 bits per channel use, indicating the severe advantage of the model given by (\ref{eq:modelBob}) and (\ref{eq:modelEve}) in increasing our ability to keep secrets. Thus we see needs to both justify the model with additional research and study additional models that make less severe assumptions. The block length of the ``best'' code is $n=4$, which is much smaller than anticipated. We believe the optimal code will grow as our analysis takes into account additional practicalities.

\section{Conclusions and Future Work}
\label{sec:conclusion}

In this paper, channel sounding measurements were taken in an indoor environment similar to what could be found in a traditional academic or industrial setting. RM wiretap codes were applied to the data and the worst-case equivocation for each was used to determine the overall best code for that specific location and setup, which turned out to be an RM(1,2) code. Careful placement and fine tuning of the transmitter power, along with correct choice of the wiretap code were all shown to be important in maximizing secure throughput over the real-world communication scenario. 

Future work will continue to both justify the channel modeling approach taken in this paper, and consider additional channel models that make fewer practical assumptions. Additional channel sounding measurements can also impart greater understanding of the distribution of $\mu$, the number of revealed bits to an eavesdropper, as a function of eavesdropper location. It is expected that this last enhancement will cause longer codes to yield higher relative levels of secure throughput when compared to shorter codes, although perhaps at an additional cost of coding rate. 


\bibliography{bibResources}
\bibliographystyle{ieeetr}

\end{document}